\documentclass[
aip,
amsmath,amssymb,
reprint,
]{revtex4-1}

\usepackage{graphicx}
\usepackage{dcolumn} 
\usepackage{bm}

\usepackage[utf8]{inputenc}
\usepackage[T1]{fontenc}
\usepackage{mathptmx}
\usepackage{etoolbox}
\usepackage{xcolor}
\usepackage{booktabs} 
\usepackage{placeins} 

\usepackage{url}
\usepackage[hyperindex,breaklinks]{hyperref}
\hypersetup{linkcolor=blue, citecolor=red, colorlinks=true}

\usepackage[margin=1.9cm]{geometry}

\makeatletter
\def\@email#1#2{%
 \endgroup
 \patchcmd{\titleblock@produce}
  {\frontmatter@RRAPformat}
  {\frontmatter@RRAPformat{\produce@RRAP{*#1\href{mailto:#2}{#2}}}\frontmatter@RRAPformat}
  {}{}
}%

\usepackage{soul}

\makeatother
\begin{document}

\title[Aperiodic Parrondo's Games]{Parrondo's effects with aperiodic protocols}

\author{Marcelo A. Pires}
\thanks{marcelo.pires@delmiro.ufal.br}
\affiliation{ 
Universidade Federal de Alagoas, Campus do Sertão, Delmiro Gouveia - AL, 57480-000, Brazil
}%

\author{Erveton P. Pinto}%
\affiliation{ 
Universidade Federal do Amapá, Macapá - AP, 68903-419, Brazil 
}%

\author{Rone N. da Silva} 
\affiliation{%
Secretaria Municipal de Educação de Gurupá,  Gurupá - PA, 68300000, Brazil 
}%

\author{S\'\i lvio M. \surname{Duarte~Queir\'{o}s}}
\thanks{sdqueiro@cbpf.br}
\affiliation{ 
Centro Brasileiro de Pesquisas F\'{\i}sicas, Rio de Janeiro - RJ, 22290-180, Brazil
}%
\affiliation{National Institute of Science and Technology for Complex Systems, Rio de Janeiro - RJ, 22290-180, Brazil}

\date{\today}

\begin{abstract}
In this work, we study the effectiveness of employing  archetypal aperiodic sequencing -- namely Fibonacci, Thue-Morse, and Rudin-Shapiro  -- on the Parrondian effect. From a capital gain perspective, our results show that these series do yield a Parrondo's Paradox with the Thue-Morse based strategy outperforming not only the other two aperiodic strategies but benchmark Parrondian games with random and periodical ($AABBAABB\ldots$) switching as well. The least performing of the three aperiodic strategies is the Rudin-Shapiro. 
To elucidate the underlying causes of these results, we analyze the cross-correlation between the capital generated by the switching protocols and that of the isolated losing games. This analysis reveals that a strong  anticorrelation  with both isolated games is typically required to achieve a robust manifestation of Parrondo's effect. 
We also study the influence of the sequencing on the capital using the lacunarity and persistence measures. In general, we observe that the switching protocols tend to become less performing in terms of the capital as one increases the persistence and thus approaches the features of an isolated losing game. For the (log-)lacunarity, a property related to heterogeneity, we notice that for small persistence (less than 0.5) the performance increases with the lacunarity with a maximum around 0.4. In respect of this, 
our work shows that the optimisation of a switching protocol is strongly dependent on a fine-tuning between persistence and heterogeneity.

\vspace{0.1cm}
{\footnotesize \url{https://pubs.aip.org/aip/cha/article-abstract/34/12/123126/3323833/Parrondo-s-effects-with-aperiodic-protocols?redirectedFrom=fulltext}}
\normalsize
\end{abstract}

\maketitle

\begin{quotation}
Common wisdom constantly tells us two harmful actions do not make a positive one. There are instances thereof in abundance. However, common wisdom often faces counter-intuitive examples too; there are particular cases -- even in Nature -- for which an optimised combination of individually negatively impacting actions can lead to a long-term positive outcome. The Parrondo's effect (also called Parrondo's Paradox) in which the combination of two losing strategies results in a winning plan of action has been employed in several fields of science and technology. These combinations of strategies are strongly inclined to present periodic arrangements, which are only a (relevant) part of non-random sequential schemes. Within this context, mathematics has provided us with aperiodic non-random sequences the applicability of which was proved in a widespread variety of systems. By intermingling the two paths, we naturally arrive to worthwhile questions over the actual impact of periodicity in switching protocols leading to a Parrondo's effect -- or, in other words -- to what extent it is possible that the combination of losing games assuming aperiodical sequencing can be Parrondian as well. 
\end{quotation}

\section{\label{sec:intro} Introduction}

From basic problems such as Archimedean hydrostatics to more sophisticated cases of special relativity, Physics and other sciences are full of counter-intuitive results\footnote{
These counter-intuitive phenomena are often called paradoxes. In this case, the term paradox is `veridical', i.e., one whose ‘proposition’ or conclusion is in fact true despite its air of absurdity and not a `falsidical' paradox\cite{quine1976ways}.
}. 
Along these lines, ratchets turned into a trendy topic in Statistical Physics and related fields\cite{feynman1963,parrondo1996criticism} when `paradoxes' are the subject, because of its relation to the capacity to perform useful work. That especially includes the case of the so-called flashing ratchet\cite{prost1994asymmetric}, where a periodic potential in space (e.g., sawtooth profile) is switched on and off. With that, one manages to impose a drift -- i.e., to perform a net useful work -- to a Brownian particle even though the potential presents an overall zero gradient.
It was in this framework that J.M.R. Parrondo devised a combination of losing games into an advantageous strategy that switches between the losing games\cite{parrond1996cheat}.

Despite the fact that similar counter-intuitive concepts were also conveyed in other subjects\cite{abbott2001overview} such as biophysics\cite{westerhoff1986enzymes}, nonlinear dynamics\cite{key1987computable}, granular matter\cite{rosato1987brazil}, random diffusion\cite{pinsky1992some}, biochemistry\cite{ajdari1992drift}, finance\cite{maslov1998optimal,luenberger1998investment}, and ecology\cite{jansen1998populations} before the publication of the Parrondo's effect itself, the model has been pivotal in helping understand how the stochastic thermal fluctuations or other sources of fluctuation and disorder -- e.g., imperfections in electronic devices\cite{bucolo2021imperfections} --   actually help achieve a positive outcome in a variety of systems ranging from exact and natural sciences to sociology and engineering\cite{rosato1987brazil,pinsky1992some,heath2002discrete,cheong2018,schwartz2009,cheong2016paradoxical,cheong2019paradoxical,wen2024parrondo}.
Because of their conceptual association with the flashing ratchet mechanism, Parrondian systems have been studied paying particular attention to the impact of the periodicity or quasi-periodicity\cite{rech2007generation} in which the two losing games, $A$ and $B$, are combined to establish regular sequences of $A$s and $B$s or purely random schemes; midway, we have aperiodic sequences\cite{lind2001distribution} that are obtained by the application of specific deterministic rules, but for which it is not possible to establish a regular pattern\cite{lynn1989}. 
Given the widespread usefulness of aperiodic sequences it is worth testing whether combinations of losing games ruled by that sort of sequencing can yield Parrondian systems. This is the main goal of the present work.

Furthermore, we introduce a novel approach by applying previously unexplored measures within the Parrondian framework. This investigation aims at elucidating the influence of structural attributes inherent to switching protocols on the manifestation of the Parrondo's effect. Specifically, we will examine the impact of lacunarity and persistence on the outcomes of the Parrondo's effect.

The remaining of the paper is organised as follows: in Sec.~\ref{sec:LitRev}, we describe the milestones in the study of Parrondo's effect, the importance of aperiodic series in science and to what extent our work compares with previous research. In Sec.~\ref{sec:method}, we introduce the methodology we employ to generate our Parrondo's games. In Sec.~\ref{sec:results}, we present our results regarding the performance in terms of capital gain of the aperiodic switching protocols and the relation between the performances of the strategy and that of the elemental isolated games as well as how structural properties of the aperiodic sequences like lacunarity and persistence define the capital attained. Last, in Sec.~\ref{sec:final}, we provide an overall picture of our results, its applications, and directions for subsequent work.

\section{\label{sec:LitRev} Literature Review}

\subsection{General aspects}

As stressed in  Ref.~\onlinecite{Abbott2009}, the original Parrondo's games were conceptualized in 1996 as a didactic representation of a flashing Brownian ratchet\cite{parrond1996cheat}. The primary Parrondo’s games\cite{abbott2010asymmetry} were established using basic coin-tossing models that lead to  paradoxical situations wherein individually losing games collectively culminate in a winning outcome. 
In 1999, Harmer and Abbott\cite{harmer1999losing} described the Parrondo's games in an explicit way. The fundamental connection between Parrondo's effect and physical phenomena has emerged as a focal point of investigation\cite{harmer2002review,parrondo2004brownian}.
Moreover, the Parrondo's effect has also been connected to  quantum games and algorithms\cite{meyer2002quantum,flitney2004quantum,chandrashekar2011parrondo,flitney2012quantum,guo2013,rajendran2018playing,pires2020parrondo,ximenes2024parrondo}. In Ref.~\onlinecite{lai2020parrondo}, the authors provide an overview of quantum Parrondo games, tracing their development from classical counterparts. 
In addition to physics, the Parrondo's effect has found connections in many other scientific domains\cite{cheong2019paradoxical,capp2021does,wen2024parrondo,molinero2023markov,gao2024analysis} including
biology\cite{macia2022aperiodic}, condensed matter physics\cite{barber2008aperiodic}, and photonics\cite{steurer2007photonic,dal2012deterministic}.
Regarding aperiodic sequences, it is worth noting that these structures have applications across multiple disciplines, including
biology\cite{macia2022aperiodic}, condensed matter physics\cite{barber2008aperiodic}, photonics\cite{steurer2007photonic,dal2012deterministic}. Furthermore, aperiodic sequences play a significant role in theoretical ecology\cite{pires2022randomness} and in the design of quantum algorithm protocols\cite{pires2020aperiodic} and spin systems\cite{pinho2019algorithm,andrade2003break}. Among aperiodic sequences\cite{barber2008aperiodic}, the Fibonacci (Fb), the Thue-Morse (TM), and the Rudin-Shapiro (RS) binary series have found their place under the limelight on its own right due to its relation to several measurable and implementable processes\cite{wahab2023,shallit1999,jagannathan2021fibonacci,baake2024fibonacci}.

\subsection{Comparison with previous related works}

In Ref.~\onlinecite{luck2019parrondo}, Luck analyses how the Parrondo's effect is impacted by the type of the switching protocol. He uses several periodic sequences as well as one aperiodic sequence, the Fb case. 

In Refs.~\onlinecite{arena2003,tang2004}, the authors considered well-known deterministic non-linear dynamics -- namely the Logistic, Tent, Sinusoidal, Gaussian, Henon, and Lozi maps -- to define the games sequences using a Genetic Algorithm and a threshold value, respectively. These authors showed that the best Parrondian strategy befalls when these chaotic generators (CG) tend to periodic behavior and that they systematically beat the random generator, especially the case based on the Logistic map. On the other hand, the CGs can yield losing strategies when the initial conditions correspond to a fixed point in the respective map, because in this case the CG is equal to the independent operation of the two losing strategies.
\\
Different from these works:
\begin{enumerate}
\item we use the 3 paradigmatic aperiodic sequences (Fb, TM and RS) that allow us to uncover the role of distinct types of aperiodicity as such sequences have different structural properties.

\item we employ a range of measures to elucidate the relation between the structural characteristics of switching protocols and the Parrondo's effect. 

\item we intrinsically reduce the number of parameters to be adjusted in the strategy in comparison to deterministic CGs\cite{arena2003,tang2004}, which can be classified as aperiodic depending on the values of its parameters. That enhances the comprehension of the role of aperiodicity in Parrondian games, since the strategies we investigate do not depend on any threshold imposed to the chaotic generator nor its attractor (established by the map parameters and the initial condition).  
\end{enumerate}

\section{\label{sec:method} Methodology}

\subsection{Game Rules}

In accordance with the established Parrondian framework \cite{harmer1999losing,harmer2002review,abbott2010asymmetry}, we define $C_t$ as the capital at time $t$. A successful game outcome results in a unit increase in capital ($C_{t+1} = C_{t} + 1$), whereas an unsuccessful outcome leads to a unit decrease ($C_{t+1} = C_{t} - 1$).
Then\cite{harmer1999losing,harmer2002review,abbott2010asymmetry}:
\begin{itemize}
    \item For game A, we use a biased coin with a probability of success defined as $P_{1} = \left(\frac{1}{2}\right) - \epsilon$.

    \item For game B, we use a conditional strategy. If the capital $C_{t}$ is an integer multiple of a constant $M$, we use a biased coin with a success probability of $P_{2} = \left(\frac{1}{10}\right) - \epsilon$. Otherwise, a distinct biased coin with a success probability of $P_{3} = \left(\frac{3}{4}\right) - \epsilon$ is employed.
\end{itemize}

\begin{figure*}
\includegraphics[scale=0.55]{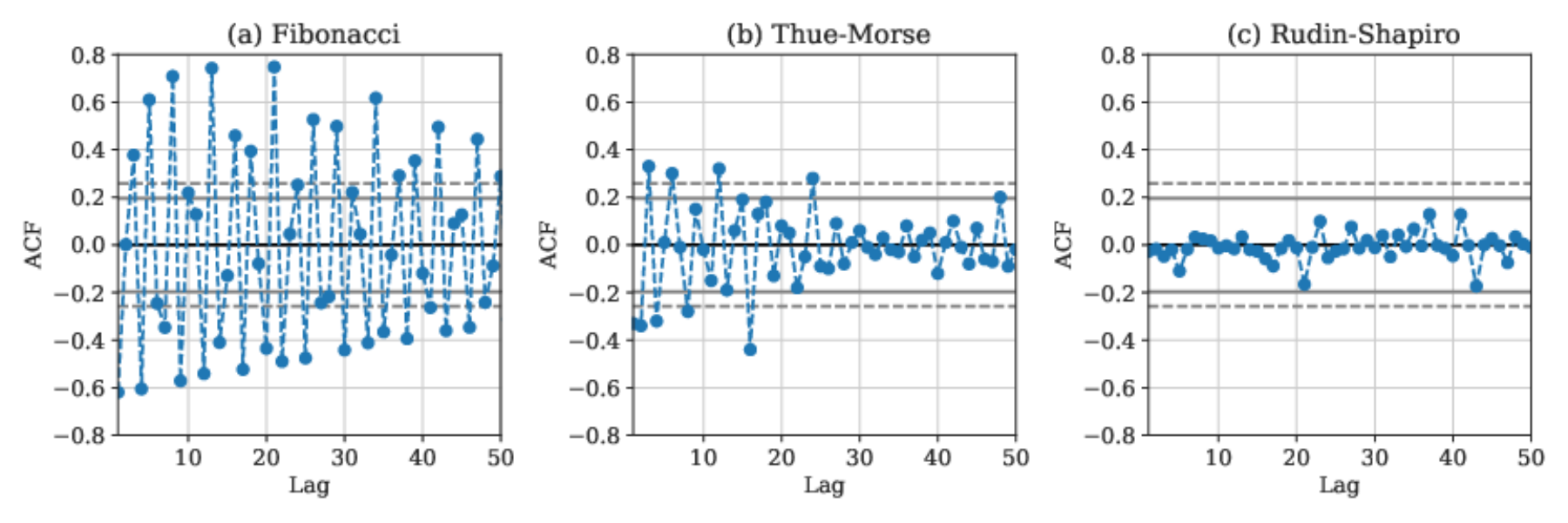}
\caption{Autocorrelation Function (ACF) for various lags for the aperiodic sequences we used (Fb, TM, RS) considering $t_{max}=100$.}
\label{fig:ACF-aperiodicseq}
\end{figure*}
\begin{figure*}
    \centering
    \includegraphics[width=0.9\linewidth]{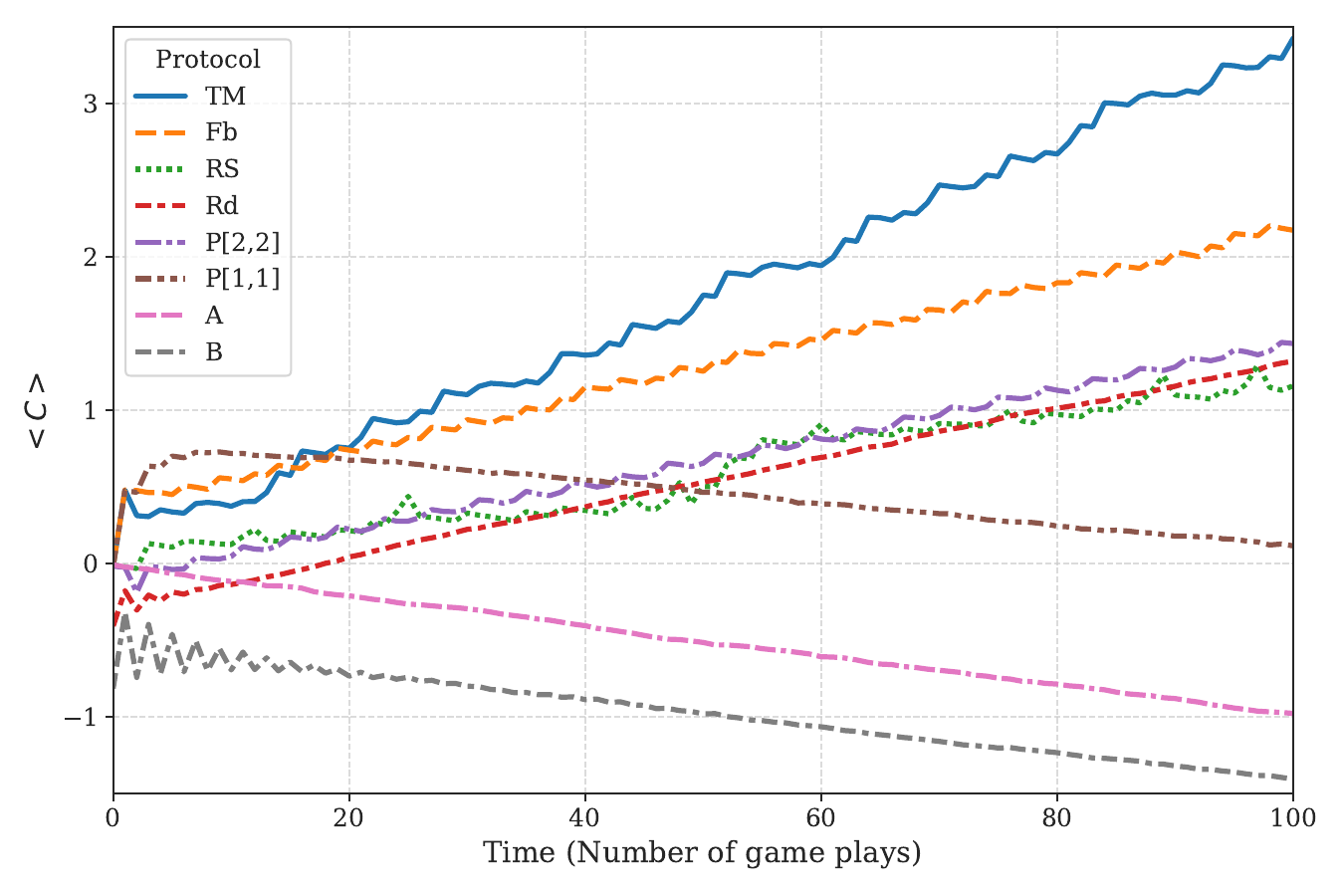}
    \caption{Mean total capital versus time for different protocol considering $t_{\max}=100$.
    }
    \label{fig:capital-vs-time-vs-protocols}    
\end{figure*}

\subsection{Aperiodic switching protocol}

Following the usual literature related to aperiodic sequences\cite{barber2008aperiodic,steurer2007photonic,dal2012deterministic,pires2022randomness,pires2020aperiodic}, we define a binary variable $b_t \in \{0, 1\}$ to regulate the protocol dynamics. The initial state is defined by $b_0 = 0$. The subsequent values of $b_t$ are generated according to one of the following rules: 
\begin{itemize}
    \item Fibonacci (Fb): Employ the substitution rules $0 \rightarrow 01$ and $1 \rightarrow 0$;
    \item Thue-Morse (TM): Utilize the substitution rules $0 \rightarrow 01$ and $1 \rightarrow 10$;
    \item Rudin-Shapiro (RS): Generate a four-letter sequence using the substitutions $A \rightarrow AB$, $B \rightarrow AC$, $C \rightarrow DB$, and $D \rightarrow DC$. Map A and B to 0, and C and D to 1.     
\end{itemize}
Next, we consider that 0 corresponds to Game A and 1 to Game B.

\begin{figure*}
    \centering
    \includegraphics[width=0.95\linewidth]{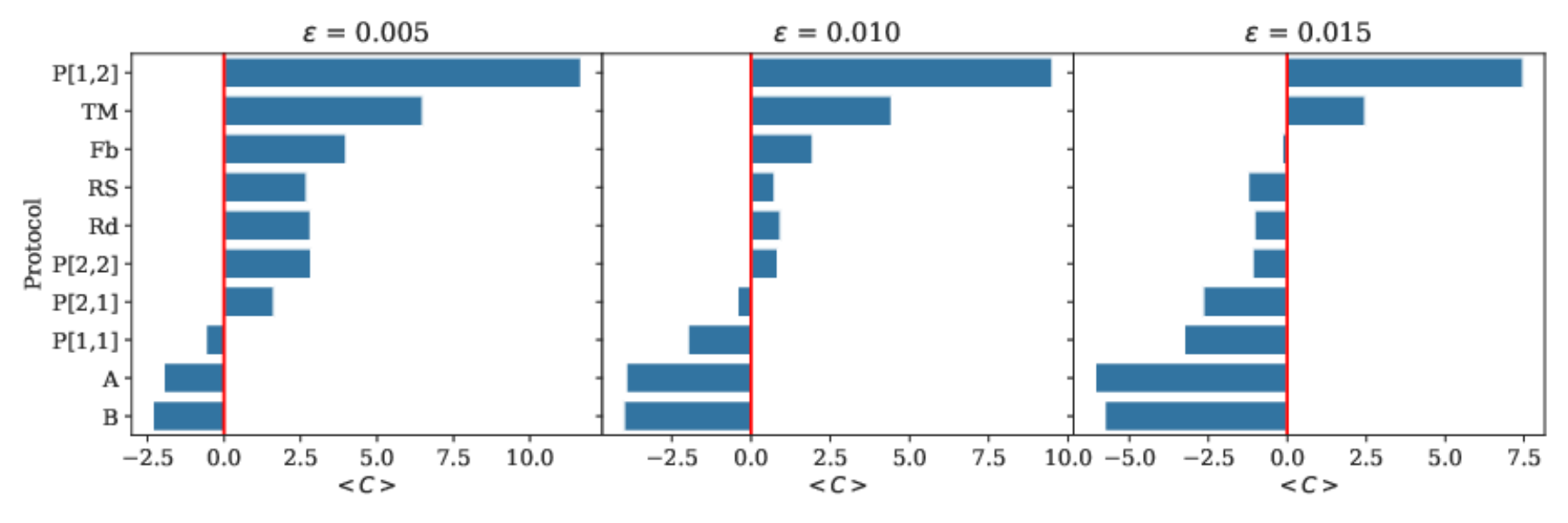}
    \caption{Barplot of the mean total capital at $t=200$ for the biasing parameter $\epsilon=\{0.005, 0.010, 0.015\}$ and different switching protocol. }
    \label{fig:barplot-capital-vs-protocol}
    \end{figure*}
\begin{figure*}
    \centering
    \includegraphics[width=0.95\linewidth]{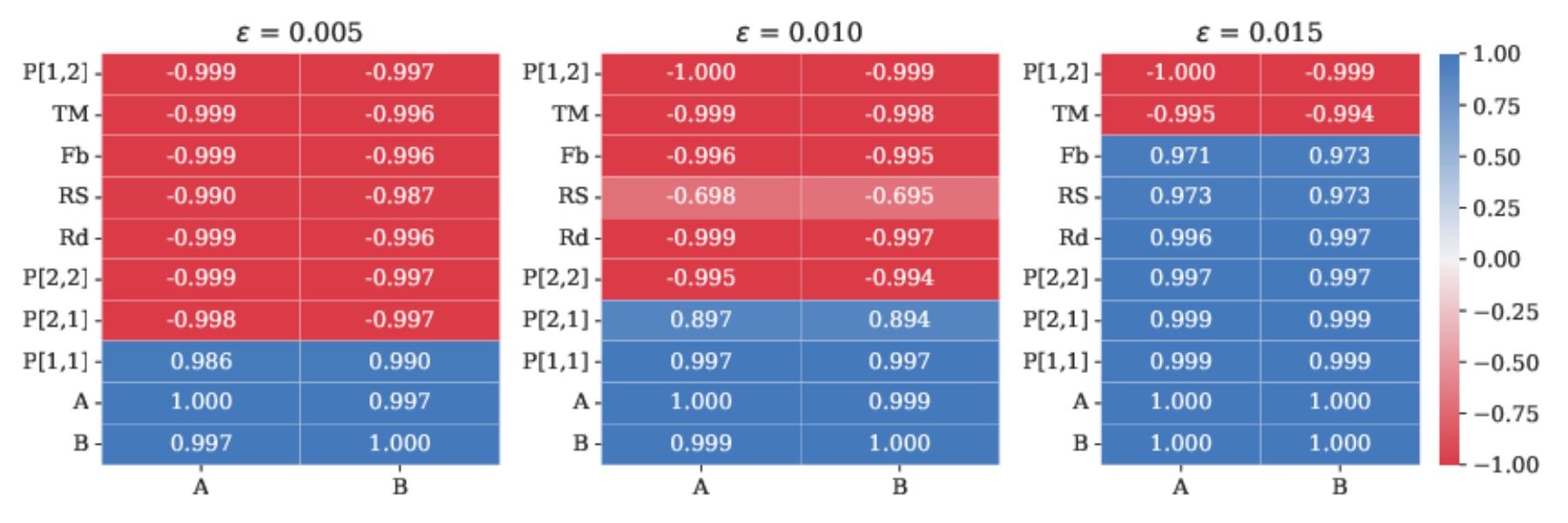}
    \caption{Pearson Cross-correlation $CC(S,X)$ between the capital of the switching protocol (S) with the game $X = \{A \ or \ B\}$. 
     Results for $\epsilon=\{0.005, 0.010, 0.015\}$ and  $t_{max}=200$.
    Similar findings are obtained if we compute $CC(S,X)$ using the Spearman and Kendall methods as shown in Appendix~\ref{app_extra_res}.}
    \label{fig:cross-correlation}
\end{figure*}

\subsection{Generalized periodic switching protocol}

To generate a periodic binary sequence composed of alternating blocks of zeros and ones, we define the following parameters: (i) $L_0$, the length of a block of zeros, (ii) $L_1$, the length of a block of ones.
Next, we grow a binary sequence until it reaches the length $L$. Then, we map $0$ to Game A and $1$ to Game B.

We introduce the notation $P[L_0, L_1]$ to concisely represent a periodic sequence where each minimal block consists of $L_0$ zeros (A) followed by $L_1$ ones (B). For instance, in Ref.~\onlinecite{harmer1999losing}, the sequence P[2,2] was used, whose initial portion is $00110011\ldots$ or $AABBAABB\ldots$.

\begin{figure*}
    \centering
    \includegraphics[width=0.90\linewidth]{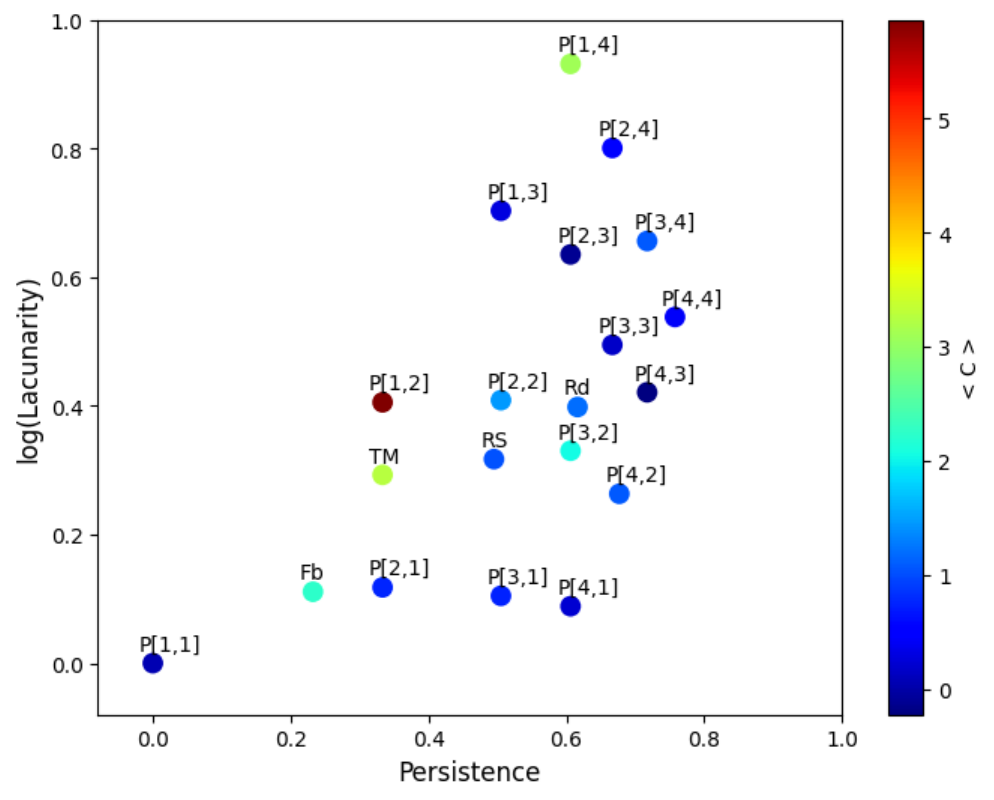}
    \caption{Mean total capital versus the Lacunarity and persistence measures of the input sequences used as a switching protocol considering $t_{max}=100$. For details on these measures see Appendix~\ref{app_lac_persis}.
    }
    \label{fig:lac-pers-plane}
\end{figure*}

\section{\label{sec:results}Results and discussion}

In this section, we present and discuss the results of two previously underexplored aspects related to Parrondian phenomena.

In Part~I, we analyze how the Parrondo's effect is impacted by distinct types of aperiodic protocols. To establish a benchmark, the traditional periodic and random switching protocols are included for comparative purposes.

In Part~II, we delve into the interplay between the structural properties of switching protocols and the Parrondo's effect, employing a variety of quantitative measures. 

Unless otherwise stated, we follow Harmer~\&~Abbott\cite{harmer1999losing} and set $\epsilon=0.005$ and $M=3$. We use $N_{\rm samples}$ from $10^4$ to $10^5$ to compute the mean capital $\langle C(t) \rangle$ at each time step $t$.

\subsection{Part I}

Figure~\ref{fig:ACF-aperiodicseq} presents the autocorrelation function (ACF) for various lags of the aperiodic sequences employed in this study (Fb, TM, RS). 
The ACF pattern of the RS sequence closely resembles that of a truly random binary sequence exhibiting a negligible serial correlation. In contrast, both the Fb and TM sequences present discernible levels of autocorrelation. However, a notable distinction emerges between these two sequences: the Fb sequence displays a considerably stronger autocorrelation compared to the TM sequence.

Figure~\ref{fig:capital-vs-time-vs-protocols} illustrates the time series of the evolution of the mean total capital for various switching protocols. 
First, note that we recover the results of Ref.~\onlinecite{harmer1999losing} regarding the periodic P[2,2] and random protocol. When considering the  capital accumulation, the strategies employing the TM and Fb sequences consistently generate superior returns compared to the strategy based on the uncorrelated RS sequence. Still with respect to capital growth, the protocol utilizing the TM sequence surpasses the performance of the protocol employing the Fb sequence. This is a surprising outcome since the TM sequence has a less pronounced autocorrelation structure than the Fb sequence.
Explicitly, the hierarchy of autocorrelation depicted in Fig.~\ref{fig:ACF-aperiodicseq} does not directly translate into the capital accumulation observed in Fig.~\ref{fig:capital-vs-time-vs-protocols}.

Figure~\ref{fig:barplot-capital-vs-protocol} presents a comprehensive overview of the mean total capital achieved for several protocol across various biasing parameters, $\epsilon$. A salient finding, is the robust performance of the TM protocol, consistently generating the highest capital accumulation, among the aperiodic protocols, for different values of $\epsilon$.  As  $\epsilon$ increases, the magnitude of the Parrondo's effect diminishes, eventually disappearing. For instance, for $\epsilon=0.015$
we observe the absence of the Parrondo's paradox in both the traditional P[2,2] and random protocols. We also observe a similar absence for the Fb and RS sequences, suggesting the aperiodic nature of these protocols does not inherently guarantee the occurrence of the Parrondo's effect.

\subsection{Part II}

Figure~\ref{fig:cross-correlation} presents the cross-correlation, $CC(S,X)$, between the capital  of the switching protocol and the underlying game $X = \{A \ or \ B\}$. 
We quantify the cross-correlation by calculating the Pearson correlation coefficients. Notably, comparable outcomes are achieved when employing Spearman or Kendall correlation coefficients, as detailed in Appendix~\ref{app_extra_res}.
The analysis encompasses a range of switching protocols, including aperiodic sequences as well as periodic and random strategies. Intriguingly, the cross-correlations between the capital and each individual game, $CC(S,A)$ and $CC(S,B)$, for the aperiodic protocols exhibit relatively small differences. This finding is somewhat unexpected considering the pronounced disparities observed in the autocorrelation functions (ACFs) of the Fb, TM, RS  sequences, as previously depicted in Fig.~\ref{fig:ACF-aperiodicseq}. While the ACF analysis revealed substantial structural differences among these sequences, their impact on the correlation between the capital and individual games appears to be less pronounced. This suggests the relation between the capital trajectory and the underlying game dynamics is more complex than simply reflecting the autocorrelation properties of the switching protocol.

Figure~\ref{fig:lac-pers-plane} presents the lacunarity and persistence measures computed for blocks of size $m=2$. This particular block size was selected due to its superior discriminatory power in differentiating between the analyzed sequences (see appendix~\ref{app_lac_persis}). 
With the exception of the highly structured P[1,1] arrangement, all evaluated sequences exhibit persistence values exceeding 0.2 and a discernible degree of heterogeneity as quantified by the lacunarity measure. It is noteworthy that sequences sharing a common level of persistence can yield significantly different mean total capital values, suggesting heterogeneity, as captured by lacunarity, plays a crucial role in determining the overall performance. However, the relation between both measures and the mean capital is not straightforwardly monotonic.
That is, the Fig.~\ref{fig:lac-pers-plane}  reveals a non-monotonic relation between the lacunarity and persistence of the switching protocol and the outcomes of the Parrondo's effect.

\section{\label{sec:final} Final remarks}
In this paper, we have explored the Parrondo's effect focussing on the role played by the aperiodicity in the performance of capital dependent Parrondian games, namely the paradigmatic Fibonacci (Fb), Thue-Morse (TM) and Rudin-Shapiro (RS) sequences. These games have been widely studied considering periodic and random combinations of losing games and to some extent considering deterministic chaotic generators, which are epistemologically the closest scenario to our analysis. 
Our results can be described twofold: 
(i) strictly looking at the performance and 
(ii) surveying the relation between the structural properties of switching protocol and the Parrondo's effect.

In respect of the former, we have verified that both the Fb and TM aperiodic generators provide outperforming games in comparison to both random and the canonical periodical game  $P[2,2]:  AABBAABB\ldots$. Additionally, the Fb generator displays a performance close to chaotic generators assuming the sinusoidal and Gaussian maps\cite{tang2004} whereas the TM generator has the best performing capital curve that is close to the Henon map\cite{arena2003}. The TM game is also robust to changes of the parameter $\epsilon $ up to $\epsilon = 0.015$ whereas all the others strategies, except P[1,2], turn into a losing switching protocol for any value of $\epsilon \le 0.015$.

The RS generator is the case in which the capital depicts the roughest curve that makes it outperform the random switching protocol during some periods and underperform the latter in others. Actually, for $t=10^4$ it is not possible to distinguish the distribution of capital between RS and the random protocol with statistical significance\footnote{The same occurs for the canonical quintessential periodical Parrondo's game\cite{harmer1999losing}}. Bearing in mind that the autocorrelation function of the RS sequence is equal to that of a white noise, this result is understandable. It is worth noting that although the white noise correlation function, the application of non-linear indicators (e.g. Lempel–Ziv measure) shows that the RS sequence is not a purely random\cite{pires2020aperiodic,pires2022randomness}.

In order to shed further light on the relation between the features of the Parrondian game and its effectiveness, we have carried out the analysis looking at the cross-correlation between the capital  of a switching protocol and its underlying games. For our aperiodic protocols, we have found the cross-correlations between the capital and each individual game are strongly negative, an unsurprising result since each isolated game is a losing strategy and the protocol combining them is a winning one. 
As a matter of fact, we expect the more  anticorrelated  the outcome of the switching protocol is with that of each isolated game, the more performing the former. This observation is best understood when we combine the cross-correlation  Fig.~\ref{fig:cross-correlation} with the capital gain results presented in Fig.~\ref{fig:lac-pers-plane}. All the three aperiodic switching protocol present very high  anticorrelation  with isolated games with TM being the most  anticorrelated  of them and the RS the very slightest less  anticorrelated  of them\footnote{For the sake of simplicity, when we mention the anti/cross-correlation between strategies and games it must be read as the anti/cross-correlation between the capital of the switching protocol and that of each isolated game.}. On the other hand, we verify that the P[1,1] (an  anticorrelated  protocol) is strongly correlated to both games, a trait that is related the poor performance of the combined game. Other periodical strategies, namely P[1,3], P[3,3], and P[4,3] go along this reasoning. This relation between cross-correlation and performance is verified for the majority of the cases, but it is neither universal nor univocal though, which makes it a future subject of study.

Still with the goal of understanding the connection between the performance of a switching protocol with the properties of its sequencing scheme, we have introduced a structural plane composed of measurements of lacunarity and persistence analogous to the complexity-entropy plane used to classify data series\cite{rosso2007distinguishing}. Overall, we observe that the switching protocols are prone to become less performing as its persistence augments. This is understood by the fact that highly persistent strategies tend to look very much like an isolated game. Yet, there is the heterogeneity assessed by (log-)lacunarity, which can dramatically change the performance. For instance, the TM switching protocol outperforms both the RS and the periodical P[3,2]. These three cases have close log-lacunarity values, but significantly larger persistence one with respect to the other; however, when we compare the performance of P[4,1] with P[1,4], which have the same persistence, we understand that the increase of lacunarity is accompanied by a significant capital hike. A close picture is learned when we go from the Fb game to the periodic P[2,1] and therefrom to the TM game or farther afield to P[1,2]. As well-known, a strategy becomes Parrondian when the second game is played at the right time leading to a positive accumulation of capital; the lacunarity is a way to gauge that sparseness with respect to one of the games (i.e., game $A$ in our case), an insight that the cross-correlation is unable to provide us with. This explains the reason game P[1,4] has high lacunarity (wherein game $A$ is less frequent) than game P[4,1], for which game $A$ is rather frequent. In future work, we expect to explore the delicate balance between lacunarity and persistence from a analytical point of view as well as surveying the performance of these aperiodic series in history-dependent switching protocols.

\begin{acknowledgments}
S.M.D.Q. thanks CNPq (Grant No. 302348/2022-0) for financial support.
\end{acknowledgments}

\begin{figure*}
    \centering
    \includegraphics[width=0.9\linewidth]{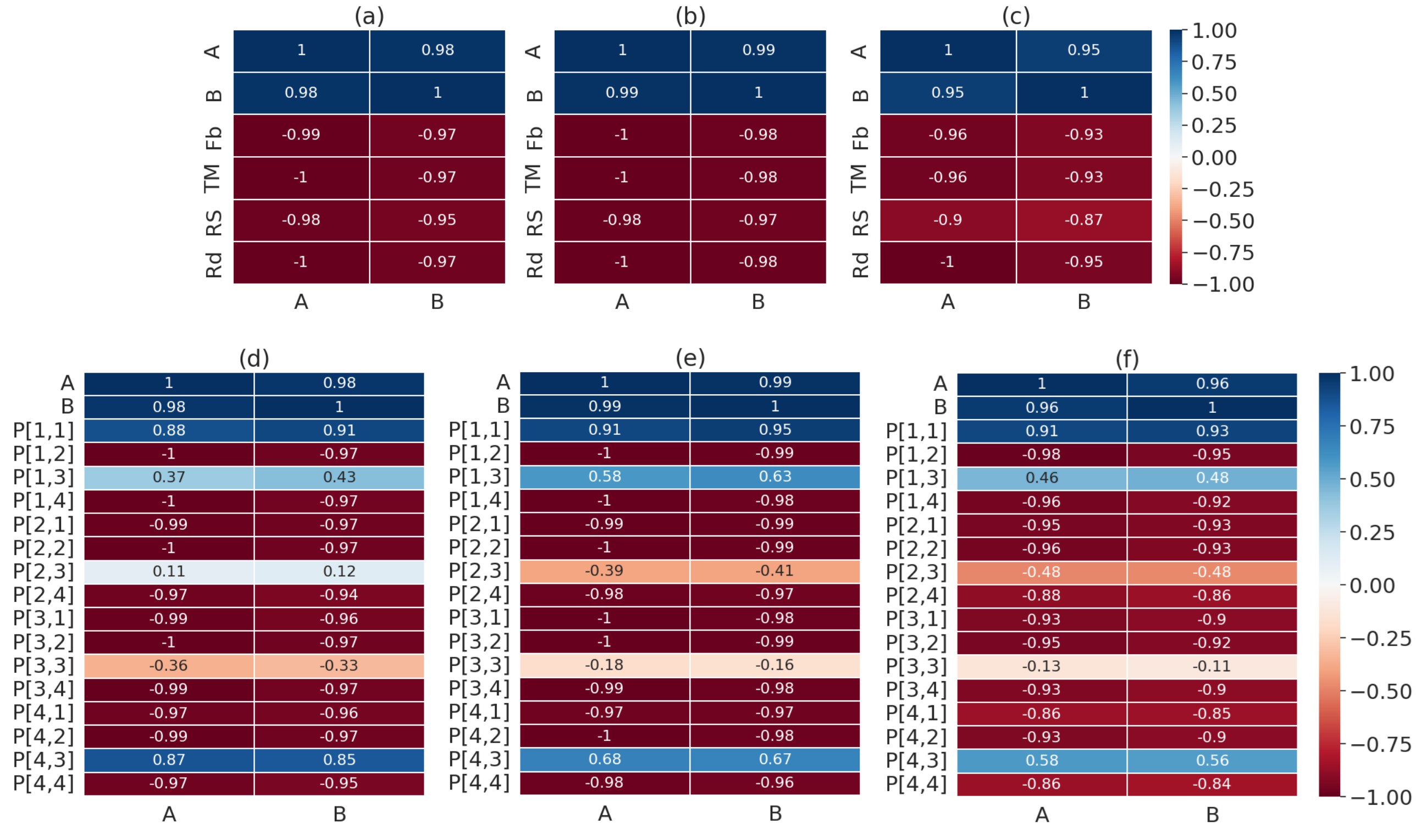}
    \caption{Cross-correlation $CC(S,X)$ between the capital of the switching protocol (S) with the game $X = \{A \ or \ B\}$ considering $t_{max}=100$.
    We use 3 methods: Pearson (a and d), Spearman  (b and e) and Kendall  (c and f). }
    \label{fig:cross-correlation-full}
\end{figure*}

\appendix

\section{Extra results}\label{app_extra_res}

Figure~\ref{fig:cross-correlation-full} shows a more comprehensive analysis of the results presented in Fig.~\ref{fig:cross-correlation} for $\epsilon = 0.005$. 
Consistent results were observed when employing Pearson's,  Spearman's and Kendall's measures of correlation, indicating a robust cross-correlation structure between the switching protocol and the selected game.
It is worth highlighting the behavior presented by the periodic arrangements with different proportions of games A and B. Based on the importance of the negative correlation for the Parrondo's effect, a greater proportion of game B does not guarantee a superior effect, nor does a greater proportion of game A guarantee an opposite behavior, even if the correlation with games A and B is strongly negative, as is the case of the arrangements P[1,2], P[2,1], P[2,2], P[1,4], P[4,1] and P[3,2], which presented very different capitals. These results reaffirm that the emergence of the Parrondo's effect is a phenomenon influenced by a multitude of factors.

\section{Lacunarity and persistence}\label{app_lac_persis}

\subsection{How to compute the lacunarity and persistence}

From the literature on lacunarity\cite{mandelbrot1995measures,plotnick1996lacunarity,allain1991characterizing,pinto2021lacunarity}, we define  $Q(m,s)$ as the probability of finding $s$ lacunas (zeros) in a block of size $m$ within a binary sequence. We also define $P(m)$, the persistence of order $m$, as the probability of observing a subsequence of length $m$ consisting entirely of zeros or ones.

\begin{figure*}
    \centering
    \includegraphics[width=0.9\linewidth]{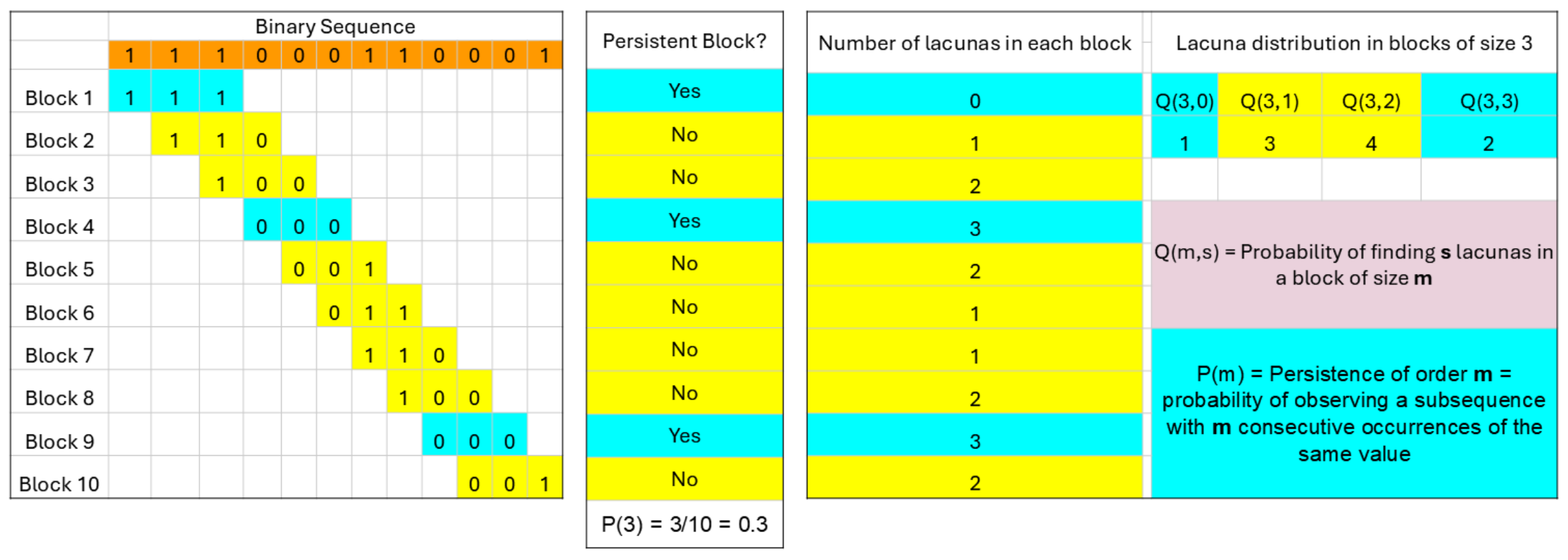}
    \caption{Steps for computing the lacunarity and persistence. As an example, we show the sequence $111000110001$.}
    \label{fig:lac-pers-per-aper-seq}
\end{figure*}

In Fig.~\ref{fig:lac-pers-per-aper-seq}, we present a concrete example to elucidate the computation of lacunarity and persistence for a binary sequence. Consider the sequence 
$111000110001$. By systematically scanning it, we determine the frequency of lacunas (zeros) and the occurrence of subsequences with identical elements. 
For example, to calculate $Q(3, s)$, we must count the number of overlapping blocks of size $m=3$ that contain exactly $s=\{0,1,2,3\}$ zeros and divide this count by $N_b$, the total number of possible blocks of size $m=3$. Similarly, to compute $P(3)$, we count the number of occurrences of $3$ consecutive ones or $3$ consecutive zeros and divide by $N_b$. See Fig.\ref{fig:lac-pers-per-aper-seq}.

\subsection{Identifying the optimal block size}

\begin{figure*}
    \centering
    \includegraphics[width=0.9\linewidth]{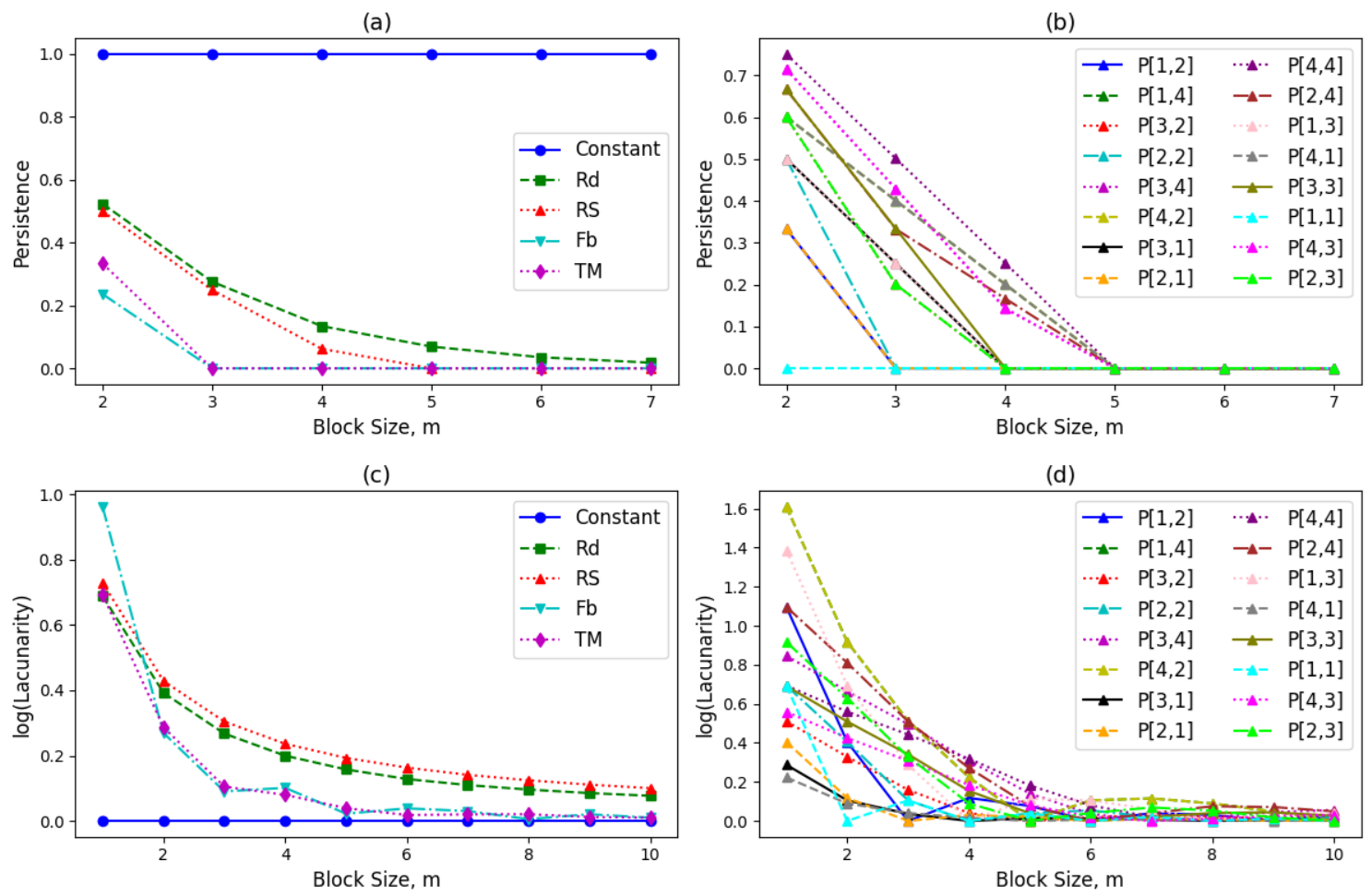}
    \caption{Persistence and lacunarity analysis for different values of m: (a) persistence for aperiodic sequences, (b) persistence for periodic sequences, (c) lacunarity for aperiodic sequences, and (d) lacunarity for periodic sequences.}
    \label{fig:lac-pers-vs-blocksize}
\end{figure*}

To effectively characterize the underlying structure of sequences through the lens of persistence and lacunarity, it is essential to determine the most suitable block size $m$. Therefore, a comprehensive analysis is conducted to evaluate the discriminatory power of different block sizes for both persistence and lacunarity. The results of this analysis are graphically presented in Fig.~\ref{fig:lac-pers-vs-blocksize}. We perceive that for the sequence types we used in this study (aperiodic and periodic) the maximum discriminatory power is observed for $m = 2$. This optimal value of $m$ is used to obtain the results shown in Fig.~\ref{fig:lac-pers-plane}.


\bibliography{main.bib}

\end{document}